\documentclass[twocolumn,showkeys,aps,prb,showpacs]{revtex4-1}
\usepackage{graphicx}
\usepackage[CJKbookmarks,dvipdfm,colorlinks,linkcolor=blue,citecolor=blue]{hyperref}

\begin{document}

\title{Intrinsic piezoelectricity in monolayer  $\mathrm{XSi_2N_4}$ (X=Ti, Zr, Hf, Cr, Mo and W)}

\author{San-Dong Guo$^{1,2}$, Yu-Tong Zhu$^{1}$ and Wen-Qi Mu$^{1}$}
\affiliation{$^1$School of Electronic Engineering, Xi'an University of Posts and Telecommunications, Xi'an 710121, China}
\affiliation{$^2$Key Laboratary of Advanced Semiconductor Devices and Materials, Xi'an University of Posts and Telecommunications, Xi'an 710121, China }
\begin{abstract}
 Motived by  experimentally synthesized $\mathrm{MoSi_2N_4}$ (\textcolor[rgb]{0.00,0.00,1.00}{Science 369, 670-674 (2020})), the intrinsic piezoelectricity in monolayer  $\mathrm{XSi_2N_4}$ (X=Ti, Zr, Hf, Cr, Mo and W)   are studied by density functional theory (DFT).  Among the six monolayers, the $\mathrm{CrSi_2N_4}$ has the best piezoelectric strain  coefficient  $d_{11}$ of 1.24 pm/V, and the second is 1.15 pm/V for $\mathrm{MoSi_2N_4}$.
 Taking $\mathrm{MoSi_2N_4}$ as a example, strain engineering is applied to improve $d_{11}$. It is found that tensile biaxial  strain can enhance
 $d_{11}$ of $\mathrm{MoSi_2N_4}$, and the $d_{11}$ at 4\% can improve by 107\% with respect to unstrained one.
By replacing the N  by P or As in $\mathrm{MoSi_2N_4}$, the $d_{11}$ can be raise substantially. For $\mathrm{MoSi_2P_4}$ and $\mathrm{MoSi_2As_4}$, the $d_{11}$ is as high as  4.93 pm/V and 6.23 pm/V, which is mainly due to smaller $C_{11}-C_{12}$ and very small minus or positive ionic contribution to piezoelectric stress  coefficient $e_{11}$  with respect to $\mathrm{MoSi_2N_4}$. The discovery of this piezoelectricity in monolayer $\mathrm{XSi_2N_4}$  enables active sensing, actuating and new electronic components for
nanoscale devices, and  is recommended for experimental exploration.

\end{abstract}
\keywords{$\mathrm{MoSi_2N_4}$, Piezoelectronics, 2D materials}

\pacs{71.20.-b, 77.65.-j, 72.15.Jf, 78.67.-n ~~~~~~~~~~~~~~~~~~~~~~~~~~~~~~~~~~~Email:sandongyuwang@163.com}

\maketitle

\section{Introduction}
Piezoelectric materials can convert mechanical energy into electrical
energy and vice versa, and  the piezoelectricity of two-dimensional (2D) materias has
been widely investigated\cite{q4} in recent years.
Experimentally, the existence of piezoelectricity of $\mathrm{MoS_2}$\cite{q5,q6}, MoSSe\cite{q8}  and $\mathrm{In_2Se_3}$\cite{q8-1}  has significantly promoted development of the piezoelectricity of 2D materials.
It has been reported that a large number of 2D
materials have significant piezoelectric coefficients, such as transition metal dichalchogenides (TMD), Janus TMD, group IIA and IIB metal oxides, group-V binary semiconductors and group III-V semiconductors\cite{q7,q7-1,q7-2,q7-3,q7-4,q9,q10,q11,q12,qr}, the  monolayer SnSe,
SnS, GeSe and GeS  of which possess   giant piezoelectricity,  as high as  75-251 pm/V\cite{q10}.
Due to different crystal symmetry,  a only in-plane piezoelectricity, both in-plane and out-of-plane piezoelectricity,  or a pure out-of-plane piezoelectricity can exit, and the corresponding example is TMD monolayers\cite{q9},  many  2D  Janus materials\cite{q7,q7-3} and penta-graphene\cite{q7-4}.
 The strain-tuned piezoelectric response of  $\mathrm{MoS_2}$\cite{r1}, AsP\cite{q7-1}, SnSe\cite{q7-1} and Janus TMD monolayers\cite{r3} have been performed by the first-principle calculations, and it is proved that strain can improve the  piezoelectric strain  coefficients.

Recently, the layered
2D $\mathrm{MoSi_2N_4}$ has been  synthesized by chemical vapor deposition (CVD)\cite{msn}. Many other 2D materials with a general formula of $\mathrm{XY_2M_4}$ have been predicted by DFT calculations\cite{msn}, where X represents an early transition metal
(W, V, Nb, Ta, Ti, Zr, Hf, or Cr), Y is Si or Ge, and M stands for N, P, or As.
In this work,  the piezoelectric properties of  monolayer  $\mathrm{XSi_2N_4}$ (X=Ti, Zr, Hf, Cr, Mo and W) are studied by using density functional perturbation theory (DFPT)\cite{pv6}.  The independent  in-plane piezoelectric constants $d_{11}$ is predicted to be 0.777 pm/V to 1.241 pm/V, which are
smaller than ones of many other 2D materials\cite{q7,q7-3,q9,q10,q11}. Using  $\mathrm{MoSi_2N_4}$ as a example, strain engineering is proposed to produce improved piezoelectric properties. It is found that increasing strain can improve $d_{11}$  due to reduced $C_{11}$-$C_{12}$ and enhanced $e_{11}$, and the band gap decreases. Calculated results show that $\mathrm{MoSi_2P_4}$ and $\mathrm{MoSi_2As_4}$ have more better $d_{11}$ than $\mathrm{XSi_2N_4}$ (X=Ti, Zr, Hf, Cr, Mo and W), which is mainly because they are more softer, and their ionic parts have very small minus contribution ($\mathrm{MoSi_2P_4}$) or positive contribution ($\mathrm{MoSi_2As_4}$) to $e_{11}$.
 Our calculations show that the $\mathrm{XY_2M_4}$ (X=Ti, Zr, Hf, Cr, Mo or W; Y=Si or
Ge; and M=N, P or As) materials may be  promising candidates for piezoelectric applications.

\begin{figure}
  \includegraphics[width=7.0cm]{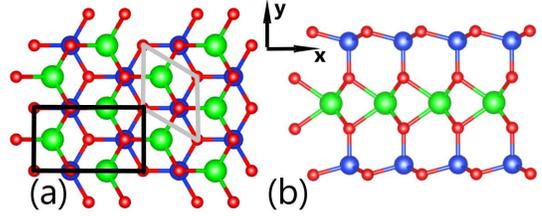}
  \caption{(Color online)The crystal structure of monolayer  $\mathrm{XSi_2N_4}$, including (a) top view and (b) side view. The primitive cell is
   are marked by gray line, and the rectangle supercell
   is marked by black line to calculate piezoelectric coefficients.  The large green balls represent X atoms, and the middle blue balls for Si atoms, and the  small red balls for N atoms.  }\label{t0}
\end{figure}


\begin{figure}
  \includegraphics[width=8cm]{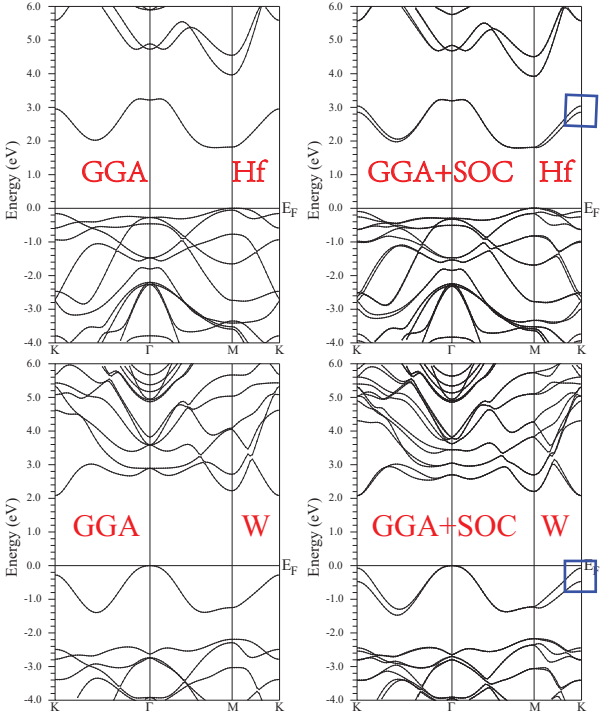}
  \caption{(Color online)The GGA and GGA+SOC energy band structures of $\mathrm{HfSi_2N_4}$ and $\mathrm{WSi_2N_4}$, and the spin-orbital splitting at K point is marked by the little blue box.}\label{band}
\end{figure}

\begin{table*}
\centering \caption{For monolayer   $\mathrm{XSi_2N_4}$ (X=Ti, Zr, Hf, Cr, Mo and W), the lattice constants $a_0$ ($\mathrm{{\AA}}$), the  height $h$ ($\mathrm{{\AA}}$), the GGA  gap $Gap$ (eV), the  GGA+SOC gap $Gap_{soc}$ (eV),  the spin-orbital splitting at K point $\Delta$ (eV),   the elastic constants $C_{11}$-$C{12}$ ($\mathrm{Nm^{-1}}$), the  piezoelectric coefficients   $e_{11}$  ($10^{-10}$ C/m ) and  $d_{11}$ (pm/V). }\label{tab0}
  \begin{tabular*}{0.96\textwidth}{@{\extracolsep{\fill}}ccccccccc}
  \hline\hline
 Name &$a_0$&$h$ &  $Gap$& $Gap_{soc}$&$\Delta$&$C_{11}$-$C_{12}$&$e_{11}$&$d_{11}$\\\hline\hline
$\mathrm{TiSi_2N_4}$ &     2.931  &   6.908    &  1.629    & 1.628    & 0.033      & 326.239  & 2.712   &   0.831                                \\\hline
$\mathrm{ZrSi_2N_4}$ &     3.032   &   7.035   &  1.629     & 1.625    & 0.056    & 287.008  & 2.229   &   0.777                             \\\hline
$\mathrm{HfSi_2N_4}$ &     3.022   &   7.000   &  1.802     & 1.789    & 0.183    & 303.898  & 3.199   &   1.053                                 \\\hline
$\mathrm{CrSi_2N_4}$ &     2.844   &   6.869   &  0.498    & 0.499    & 0.063    & 358.021  & 4.442   &   1.241                                   \\\hline
$\mathrm{MoSi_2N_4}$ &     2.909   &   7.004   &  1.747    & 1.746    & 0.130    & 383.982  & 4.398    &   1.145                                      \\\hline
$\mathrm{WSi_2N_4}$  &      2.912  &    7.014  &  2.083    & 2.074    & 0.399    & 403.227  & 3.138   &   0.778                                                                        \\\hline\hline
\end{tabular*}
\end{table*}

\begin{figure}
  \includegraphics[width=7cm]{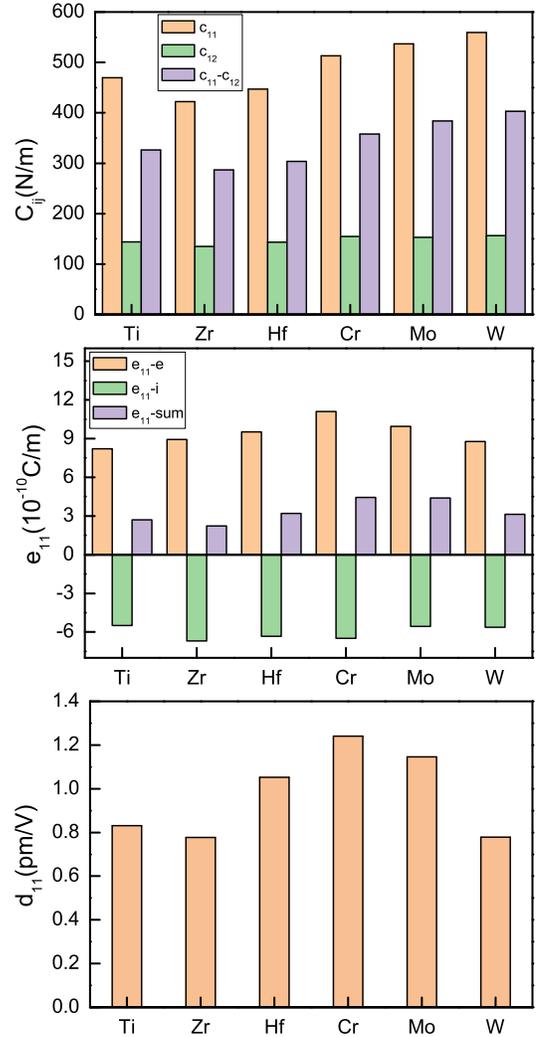}
  \caption{(Color online) For $\mathrm{XSi_2N_4}$ (X=Ti, Zr, Hf, Cr, Mo and W): (Top) the elastic constants  $C_{ij}$, (Middle) piezoelectric stress coefficients  $e_{11}$ and  the ionic contribution and electronic contribution to $e_{11}$, and (Bottom) piezoelectric strain coefficients  $d_{11}$.  }\label{cn}
\end{figure}

\section{Computational detail}
We perform DFT calculations\cite{1} using the projector-augmented wave method as implemented
in the plane-wave code VASP\cite{pv1,pv2,pv3}. For the structural relaxation and the calculations of the elastic and
piezoelectric tensors,   we use the popular  generalized gradient
approximation of Perdew, Burke and  Ernzerhof  (GGA-PBE)\cite{pbe} as the exchange-correlation  functional. For energy band calculations, the spin orbital coupling (SOC)
is also taken into account.
 A cutoff energy of 500 eV for the
plane wave basis set is used to ensure an
accurate DFT calculations.
  A vacuum spacing of more than 32 $\mathrm{{\AA}}$ is adopted to reduce the interactions between
the layers, which is key to attain accurate $e_{ij}$.
 The total energy  convergence criterion is set
to $10^{-8}$ eV, and  the Hellmann-Feynman forces  on each atom are less than 0.0001 $\mathrm{eV.{\AA}^{-1}}$.
The coefficients of the elastic stiffness tensor  $C_{ij}$   are calculated by using strain-stress relationship (SSR),  and   the piezoelectric stress coefficients $e_{ij}$ are attained by DFPT method\cite{pv6}.
 The Brillouin zone sampling
is done using a Monkhorst-Pack mesh of 15$\times$15$\times$1  for $C_{ij}$, and  9$\times$16$\times$1 for $e_{ij}$.
The 2D elastic coefficients $C^{2D}_{ij}$
 and   piezoelectric stress coefficients $e^{2D}_{ij}$
have been renormalized by the length of unit cell along z direction ($Lz$):  $C^{2D}_{ij}$=$Lz$$C^{3D}_{ij}$ and $e^{2D}_{ij}$=$Lz$$e^{3D}_{ij}$.

  \begin{figure}
  \includegraphics[width=7cm]{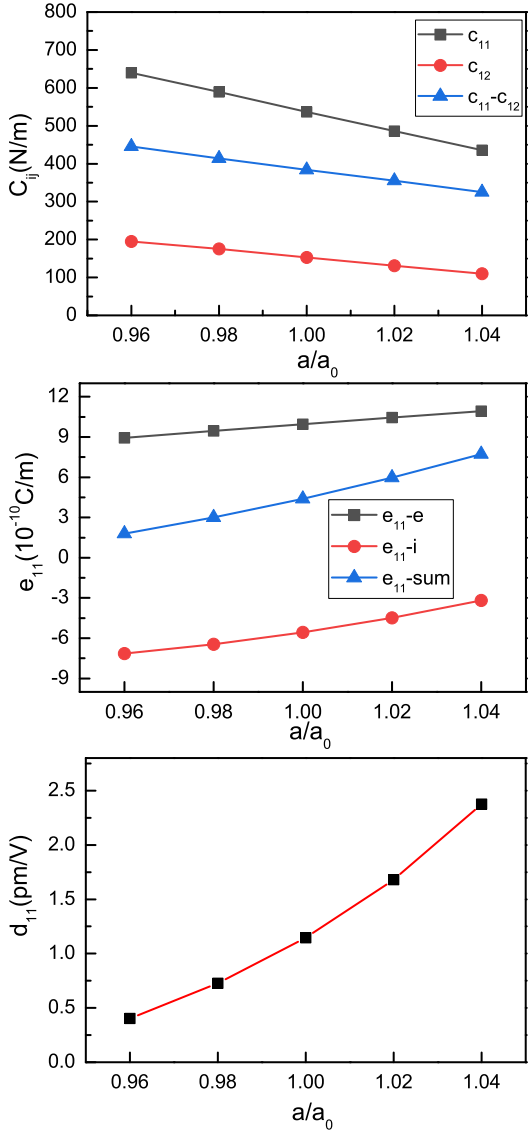}
  \caption{(Color online) For experimentally achieved monolayer $\mathrm{MoSi_2N_4}$,(Top) the elastic constants  $C_{ij}$, (Middle) piezoelectric stress coefficients  $e_{11}$ and  the ionic contribution and electronic contribution to $e_{11}$, and (Bottom) piezoelectric strain coefficients  $d_{11}$   as a function of  biaxial  strain.  }\label{cn1}
\end{figure}

\section{Symmetry Analysis}
 The relaxed-ion piezoelectric  stress tensors  $e_{ijk}$ and strain tensor $d_{ijk}$,  from the sum of ionic
and electronic contributions , is defined as:
 \begin{equation}\label{pe0}
      e_{ijk}=\frac{\partial P_i}{\partial \varepsilon_{jk}}=e_{ijk}^{elc}+e_{ijk}^{ion}
 \end{equation}
and
 \begin{equation}\label{pe0-1}
   d_{ijk}=\frac{\partial P_i}{\partial \sigma_{jk}}=d_{ijk}^{elc}+d_{ijk}^{ion}
 \end{equation}
where  $P_i$, $\varepsilon_{jk}$ and $\sigma_{jk}$ are polarization vector, strain and stress, respectively.
The $d_{ijk}$ and $e_{ijk}$ are related via the elastic stiffness tensor  $C_{ijkl}$. Monolayer  $\mathrm{XSi_2N_4}$  belongs to the $\bar{6}m2$ point group. Employing the Voigt notation, if we only consider in-plane strain components\cite{q7,q9,q10,q11,q12} for 2D materials,
 the  $e_{ij}$,  $d_{ij}$ and $C_{ij}$ become into:
 \begin{equation}\label{pe1}
  \left(
    \begin{array}{ccc}
      e_{11} &-e_{11} & 0 \\
    0 &0 & -e_{11}\\
      0 & 0 & 0 \\
    \end{array}
  \right)
  \end{equation}
  \begin{equation}\label{pe1}
  \left(
    \begin{array}{ccc}
        d_{11} & -d_{11} & 0 \\
    0 &0 & -2d_{11} \\
      0 & 0 & 0 \\
    \end{array}
  \right)
  \end{equation}
  \begin{equation}\label{pe1}
    \left(
    \begin{array}{ccc}
      C_{11} & C_{12} &0 \\
     C_{12} & C_{11} &0 \\
     0 & 0 & \frac{C_{11}-C_{12}}{2} \\
    \end{array}
  \right)
   \end{equation}
Here, the only in-plane $d_{11}$ is derived by  $e_{ik}$=$d_{ij}C_{jk}$:
\begin{equation}\label{pe2-7}
    d_{11}=\frac{e_{11}}{C_{11}-C_{12}}
\end{equation}

\section{Main calculated results}
The geometric structures of the  $\mathrm{XSi_2N_4}$ monolayer are
plotted in \autoref{t0}, which consist of  seven atomic layers of N-Si-N-X-N-Si-N (a $\mathrm{XN_2}$ layer sandwiched between two Si-N bilayers).
The optimized
structural parameters of $\mathrm{XSi_2N_4}$ (X=Ti, Zr, Hf, Cr, Mo and W) (in \autoref{tab0})  agree well with the
previous calculated results\cite{msn}.  The electronic band
structures of these  monolayers are also calculated using GGA and GGA+SOC, and the representative $\mathrm{HfSi_2N_4}$ and  $\mathrm{WSi_2N_4}$ monolayers are  shown
in \autoref{band}. The energy bands of $\mathrm{XSi_2N_4}$ (X=Ti, Zr,  Cr, Mo) are plotted in Fig.1 and Fig.2 of supplementary materials.  Compared to $\mathrm{XSi_2N_4}$ (X=Ti, Zr, Hf), additional two electrons are added for  $\mathrm{XSi_2N_4}$ (X=Cr, Mo and W), and then  the first two conduction bands are filled.
Their corresponding
 gaps (GGA and GGA+SOC) and spin-orbital splitting at K point  are summarized in \autoref{tab0}. It is clearly seen that the difference of gap between GGA and GGA+SOC is very little. Calculated results show that the magnitude of  spin-orbital splitting accords with the atomic mass of X.

Due to  hexagonal symmetry, the two independent elastic stiffness coefficients  ($C_{11}$ and $C_{12}$) are calculated by SSR, and all calculated elastic coefficients satisfy the Born stability criteria\cite{ela}.  The elastic stiffness coefficients  ($C_{11}$, $C_{12}$ and $C_{11}$-$C_{12}$) are show in \autoref{cn}. These elastic constants are larger than ones of most 2D materials, like TMD,  metal oxides, and III-V
semiconductor materials\cite{q9,q11},  indicating that these 2D
 monolayers are more  rigid than other 2D materials. The  piezoelectric stress coefficients $e_{11}$   of  $\mathrm{XSi_2N_4}$  monolayer are calculated by DFPT, using the  rectangle supercell. Based on \autoref{pe2-7}, the  piezoelectric strain coefficients $d_{11}$ are attained.
  The   piezoelectric coefficients  $e_{11}$ and  $d_{11}$, and the ionic contribution and electronic contribution to $e_{11}$  are  plotted in \autoref{cn}. Some key data are also listed in \autoref{tab0}. For all six monolayers, it is clearly seen that the ionic contribution and electronic contribution to $e_{11}$ is opposite. The entire range of
calculated $e_{11}$ is  from 2.229 $10^{-10}$ C/m to 4.442 $10^{-10}$ C/m, while the $d_{11}$  ranges from 0.777 pm/V to 1.241 pm/V. Their $d_{11}$
 are smaller than ones of TMD monolayers (2.12 pm/V to 13.45 pm/V)\cite{q9,q11}.
 For example,  the $e_{11}$ of $\mathrm{CrSi_2N_4}$ (4.442 $10^{-10}$ C/m) and $\mathrm{MoSi_2N_4}$ (4.398 $10^{-10}$ C/m) are larger than one of $\mathrm{MoS_2}$ (3.64 $10^{-10}$ C/m), but their $d_{11}$ (1.241 pm/V and 1.145 pm/V) are smaller than one of $\mathrm{MoS_2}$ (3.73 pm/V)\cite{q9,q11}, which is due to larger $C_{11}$-$C_{12}$. Among all studied six monolayers, the  $\mathrm{CrSi_2N_4}$ monolayer has the best $d_{11}$.
  \begin{figure}
  \includegraphics[width=8cm]{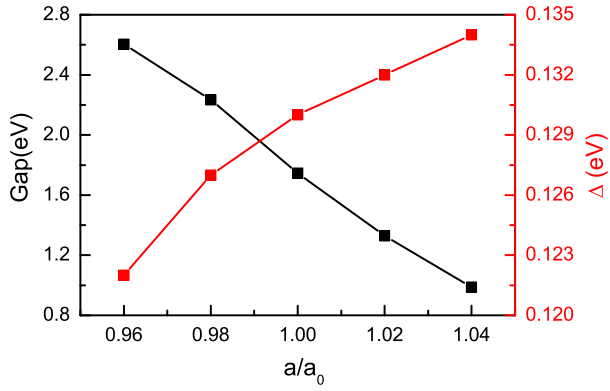}
  \caption{(Color online) For experimentally achieved monolayer $\mathrm{MoSi_2N_4}$, the GGA+SOC gap and spin-orbital splitting at K point as a function of  biaxial  strain.  }\label{gap}
\end{figure}

The $d_{11}$ of  $\mathrm{XSi_2N_4}$ monolayer is very small,  and strain engineering is proposed to enhance their piezoelectric properties, which has been proved to a very effective way\cite{r1,q7-1,r3}.
Here, we use  experimentally synthesized $\mathrm{MoSi_2N_4}$ as an example to study the strain effects on piezoelectric properties.
Due to $\bar{6}m2$ symmetry, biaxial  strain can not induce polarization, not like uniaxial  strain. We only consider biaxial  strain effects on piezoelectric properties of $\mathrm{MoSi_2N_4}$,  and the elastic constants  $C_{11}$-$C_{12}$,  piezoelectric coefficients  $e_{11}$ and  $d_{11}$, and the ionic contribution and electronic contribution to $e_{11}$ of monolayer $\mathrm{MoSi_2N_4}$ as a function of  biaxial  strain are plotted in \autoref{cn1}. When the strain varies from -4\% to 4\%, the  $C_{11}$-$C_{12}$ decreases, and the $e_{11}$ increases, which gives rise to improved
$d_{11}$  based on \autoref{pe2-7}.  At 4\% strain, the $d_{11}$ is  2.375 pm/V, which is more than twice as large as  unstrained one  (1.145 pm/V).
Similar  biaxial  strain-improved $d_{11}$ can be found in monolayer  g-$\mathrm{C_3N_4}$ and $\mathrm{MoS_2}$\cite{gsd}. It is found that both ionic contribution and electronic contribution to $e_{11}$ have positive influence to improve $d_{11}$ of monolayer $\mathrm{MoSi_2N_4}$, which is different from monolayer  g-$\mathrm{C_3N_4}$ and $\mathrm{MoS_2}$\cite{gsd}.

\begin{figure}
  \includegraphics[width=7cm]{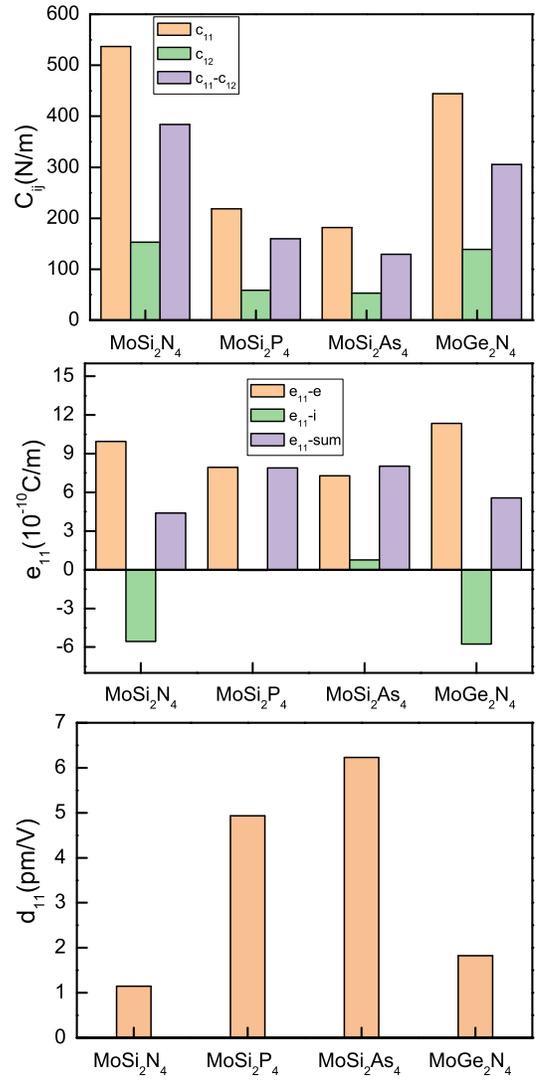}
  \caption{(Color online) For monolayer $\mathrm{MoSi_2N_4}$, $\mathrm{MoSi_2P_4}$, $\mathrm{MoSi_2As_4}$ and $\mathrm{MoGe_2N_4}$: (Top) the elastic constants  $C_{ij}$, (Middle) piezoelectric stress coefficients  $e_{11}$ and  the ionic contribution and electronic contribution to $e_{11}$, and (Bottom) piezoelectric strain coefficients  $d_{11}$.  }\label{cn3}
\end{figure}
At applied strain, the monolayer $\mathrm{MoSi_2N_4}$  exhibits piezoelectricity,  which  should have a band gap.
The gap  and  spin-orbital splitting  $\Delta$ at K point as a function of strain are  plotted in \autoref{gap}, and the strain-related energy bands of $\mathrm{MoSi_2N_4}$  are plotted in Fig.3 of supplementary materials.
It is found  that the gap decreases from 2.605 eV (-4\%) to 0.988 eV (4\%), while the $\Delta$ increases from 0.122
 eV to 0.134 eV. The position of  conduction band
minimum (CBM) do not change from -4\% to 4\%, but the position of  valence band maximum (VBM) changes from K point to $\Gamma$ point.
The valence bands convergence  can be observed at about -2\% strain due to almost the same energy between K point and $\Gamma$ point,  which is in favour of better p-type Seebeck coefficient. Similar strain-induced bands convergence can be observed in many 2D materials like $\mathrm{PtSe_2}$\cite{gsd1}.

To further enhance  piezoelectric properties,  using elements of group IVA and elements
of group VA to replace the
Si and N elements in experimentally synthesized $\mathrm{MoSi_2N_4}$,   the monolayer  $\mathrm{MoSi_2P4}$, $\mathrm{MoSi_2As_4}$ and $\mathrm{MoGe_2N_4}$ are proved to be stable\cite{msn}. The elastic constants  $C_{11}$-$C_{12}$,  piezoelectric coefficients  $e_{11}$ and  $d_{11}$, and the ionic contribution and electronic contribution to $e_{11}$ of  monolayer $\mathrm{MoSi_2N_4}$, $\mathrm{MoSi_2P_4}$, $\mathrm{MoSi_2As_4}$ and $\mathrm{MoGe_2N_4}$ are plotted in \autoref{cn3}. It is clearly seen that monolayer $\mathrm{MoSi_2P_4}$ and $\mathrm{MoSi_2As_4}$ have very higher $d_{11}$ than $\mathrm{MoSi_2N_4}$, and they are 4.93 pm/V and 6.23 pm/V, which are  comparable to one of most TMD monolayers\cite{q9}.
One reason of the high $d_{11}$ for monolayer $\mathrm{MoSi_2P_4}$ and $\mathrm{MoSi_2As_4}$ is that monolayer $\mathrm{MoSi_2P_4}$ and $\mathrm{MoSi_2As_4}$ have more smaller $C_{11}$ and $C_{12}$ than $\mathrm{MoSi_2N_4}$, which leads to smaller $C_{11}$-$C_{12}$. Another reason is that the minus of the ionic contribution  to $e_{11}$ of  monolayer $\mathrm{MoSi_2P_4}$ is very small, and the  ionic contribution is positive for monolayer $\mathrm{MoSi_2As_4}$.
The $d_{11}$ of monolayer $\mathrm{MoGe_2N_4}$  is 1.83 pm/V. which is close to one of $\mathrm{MoSi_2N_4}$.

\section{Conclusion}
Significant progress has been achieved in  synthetizing monolayer $\mathrm{MoSi_2N_4}$ with a non-centrosymmetric structure, which  allows  it
to be piezoelectric. Here, the piezoelectric properties of monolayer  $\mathrm{XSi_2N_4}$ (X=Ti, Zr, Hf, Cr, Mo and W) are studied by using first-principles calculations. In the considered six materials, the $\mathrm{CrSi_2N_4}$ is predicted to have the best  $d_{11}$ of 1.24 pm/V, and the second is 1.15 pm/V for experimentally synthesized $\mathrm{MoSi_2N_4}$. It is found that strain engineering can  improve $d_{11}$  of $\mathrm{MoSi_2N_4}$, and the $d_{11}$ at 4\% biaxial  strain can improve by 107\%. Compared to  monolayer  $\mathrm{XSi_2N_4}$ (X=Ti, Zr, Hf, Cr, Mo and W), the  monolayer  $\mathrm{MoSi_2P_4}$, $\mathrm{MoSi_2As_4}$ and $\mathrm{MoGe_2N_4}$ have more higher $d_{11}$, and the $d_{11}$ of $\mathrm{MoSi_2As_4}$ is as high as 6.23 pm/V. Owing to the recent CVD growth in  monolayer $\mathrm{MoSi_2N_4}$, it is
expected that these monolayers $\mathrm{XY_2M_4}$ (X=Ti, Zr, Hf, Cr, Mo or W; Y=Si or
Ge; and M=N, P or As) may be put to a wide practical piezoelectric use in the future.

\begin{acknowledgments}
This work is supported by the Natural Science Foundation of Shaanxi Provincial Department of Education (19JK0809). We are grateful to the Advanced Analysis and Computation Center of China University of Mining and Technology (CUMT) for the award of CPU hours and WIEN2k/VASP software to accomplish this work.
\end{acknowledgments}

\end{document}